# Field-tuned Collapse of an Orbital Ordered and Spin-polarized State: Colossal Magnetoresistance in Bilayered Ruthenate

G. Cao[1], L. Balicas[2], X. N. Lin[1], S. Chikara[1], V. Duairaj[1], E. Elhami[1], J.W. Brill[1], and R.C. Rai[1]

[1]Department of Physics and Astronomy, University of Kentucky, Lexington, KY 40506

[2] National High Magnetic Field Laboratory, Tallahassee, FL 32310

$Ca_3Ru_2O_7$ with a Mott-like transition at 48 K features different in-plane anisotropies of magnetization and magnetoresistance. Applying magnetic field along the magnetic easy-axis precipitates a spin-polarized state via a first-order metamagnetic transition, but does not lead to a full suppression of the Mott state, whereas applying magnetic field along the magnetic hard axis does, causing a resistivity reduction of three orders of magnitude. The colossal magnetoresistivity is attributed to the collapse of the orbital ordered and spin-polarized state. Evidence for a density wave is also presented.

Physics of magnetoresistance has generated enormous interest in recent years. While this quantum mechanical phenomenon is in general associated with spin scattering process of conduction electrons, the origin of various kinds of magnetoresistance is vastly different. Giant magnetoresistance (GMR) observed in magnetic metallic multiplayer structures can be qualitatively explained using the two current model, corresponding to the up-spin and down-spin electrons [1]. Tunneling magnetoresistance (TMR), often seen in magnetic tunnel junctions separated by an insulating spacer layer, is a consequence of spin-polarization. On the other hand, colossal magnetoresistance (CMR), seen only in the mixed-valence manganites so far, originates from a metal-insulator transition in the vicinity of the Curie temperature driven primarily by double exchange due to the hopping of $e_g$ electrons of $Mn^{3+}$ ions and the Jahn-Teller effect [2].

The novelty of the bilayered $Ca_3Ru_2O_7$, as presented in this letter, is that *the colossal magnetoresistivity is a result of the collapse of the orbital ordered state that is realized by demolishing the spin-polarized state.* This new phenomenon is striking in that the spin-polarization, which is a fundamental driving force for all other magnetoresistive systems, is detrimental to the colossal magnetoresistence in this 4d-based electron system! Indeed, the electron kinetic energy hinges on the spin-orbital-lattice coupling in such a way that applying magnetic field, B, along the magnetic easy-axis (*a*-axis) precipitates a spin-polarized state via a first-order metamagnetic transition, but does not lead to a full suppression of the Mott state, whereas applying B along the magnetic hard axis (*b*-axis) does, giving rise to a resistivity reduction of three orders of magnitude. Our previous work indicated the anisotropic behavior [3], but it did not establish the existence and consequences of the orbital ordering. The results presented here including temperature and field dependence of resistivity with B||*a*- and *b*-axis and nonlinear transport represent a systematic study addressing the role of the orbital ordering. It is now clear that the orbital and spin degrees of freedom are intimately coupled so that polarizing spins stabilizes the orbital ordered state that forbids the electron hopping, and thus the destruction of the spin-polarized state naturally leads to a "quantum melting" of the orbital ordered state that in turn brings about the coloassal magnetoresistance. This conclusion is also supported by evidence for the formation of a density wave along the *a*-axis.

The bilayered $Ca_3Ru_2O_7$ belongs to the Ruddlesden-Popper series with lattice parameters of *a*=5.3720(6) Å, *b*=5.5305(6) Å, and *c*=19.572(2) Å [4]. The crystal structure is severely distorted by tiling of $RuO_6$. The tilt projects primarily onto the *ac*-plane (153.22$^o$) while only slightly impacting the *bc*-plane (172.0$^o$) [4]. These are crucial bond angles defining anisotropic spin-orbital-lattice coupling within the basal plane. Recent structural studies also reveal a simultaneous collapse of the *c*-axis lattice parameter at the Mott-like transition, suggesting the importance of magnetoelastic interactions and a Jahn-Teller distortion of $RuO_6$ [3,5].

$Ca_3Ru_2O_7$ undergoes an antiferromagnetic (AFM) ordering at $T_N$=56 K while remaining metallic [6] and then undergoes a Mott-like transition at $T_{MI}$=48 K [3-11] with the formation of a charge gap of 0.1 eV that bears a resemblance to a Mott system [7, 10]. Shown in Fig.1 is the temperature dependence of magnetization, M, for the *a*- and *b*-axis for a few representative magnetic fields, B. In Fig.1a, M(T) for the *a*-axis, the magnetic easy axis, features two phase transition temperatures, $T_N$=56 K and $T_{MI}$=48 K (also see the inset) [3-11]. The low field M decreases precipitously at $T_{MI}$, indicating a magnetoelastic effect occurring upon the collapse of the *c*-axis lattice parameter mentioned above [3,5]. As B increases, $T_{MI}$ shifts slightly, whereas $T_N$ remains essentially unchanged initially and becomes rounded eventually. At B>6 T, the first-order metamagnetic transition leads to a *spin-polarized or ferromagnetic* (FM) state with a saturation moment $M_s$ =1.73 $\mu_B$/Ru, suggesting more than 85% spin polarization, assuming 2 $\mu_B$/Ru expected for an S=1 system. On the other hand, M(T) for the *b*-axis, as shown in Fig.1b and the inset, exhibits no anomaly corresponding to $T_{MI}$ but a sharp peak at $T_N$ that decreases at approximately a rate of 2K/T. Markedly, the magnetic ground state for B||*b*-axis *remains antiferromagnetic*, entirely different from that for B||*a*-axis. This anisotropic behavior reflects the anisotropic spin-orbit coupling that only favors the spin-stabilized orbital ordering at $T_{MI}$ along the *a*-axis. The absence of the anomaly at $T_{MI}$ in the *b*-axis M confirms such a spin-orbit coupling. As can be seen below, applying B along the *b*-axis destabilizes the spin-polarized state, and consequently causes the collapse of the orbital ordered state, which in turn drastically increases the electron mobility.

Shown in Fig.2 is the temperature dependence of the c-axis resistivity, $\rho_c$, at a few representative B, which is applied along the *a*-axis (Fig.2a) and the *b*-axis (Fig.2b), respectively (Note that the same log scale is used for both Figs.2a and 2b for comparison). For B||*a*-axis, $\rho_c$ at low temperatures increases slightly with increasing B when B<6 T, and decreases abruptly by about an order of magnitude when B≥6 T, at which the first order metamagnetic transition lead to the spin-polarized state (also see Fig3a). Further increasing B only results in slightly higher resistivity at low temperatures. The reduction of $\rho_c$ is attributed to a tunneling effect facilitated by a field-induced coherent motion of spin-polarized electrons between Ru-O planes separated by insulating (I) Ca-O planes. This situation is similar to an array of FM/I/FM junctions where the probability of tunneling and thus electronic conductivity depend on the angle between spins of adjacent ferromagnets. It is astonishing that the spin-polarized state generates no fully metallic state in spite of the pronounced impact of the tunneling effect on $\rho_c$. In fact, the behavior of $\rho_c$ at B=28 T suggests a still nonmetallic state (see the inset)! What is entirely unexpected is that when B||*b*-axis, the magnetic hard axis, the Mott state starts to collapse at approximately the rate of 2K/T, and vanishes at B>24 T, as clearly shown in Fig1.b.

Such an anisotropy is further illustrated in Fig.3 where $\rho_c$ as a function of B is plotted for B||*a*-axis (Fig.3a), and B||*b*-axis (Fig.3b), respectively, for a few representative temperatures. M at 5 K for the *a*-axis and the *b*-axis is also shown in Fig.3a (right scale) for the discussion. When B||*a*-axis, $\rho_c$ shows a first order transition in the vicinity of 6 T apparently driven by the first order metamagnetic transition which leads to the spin-polarized state with a saturation moment, $M_s$=1.73 $\mu_B$/Ru. This metamagnetic transition decreases slightly with increasing temperature and disappears completely when T approaches $T_{MI}$ (=48 K) [6]. Further increasing B to 30 T only results in a linear increase of the resistivity with B. When B||*b*-axis, the magnetic state remains antiferromagnetic as shown in M (*b*-axis), completely different from M(*a*-axis) due to a strong anisotropy field of 22.4 T [11] (see the inset), and $\rho_c$ rapidly decreases at a critical field by as much as three orders of magnitude shown in Fig.3b. The temperature and field dependence of the *a*-axis resistivity $\rho_a$ is nearly identical to that of $\rho_c$ presented here, therefore not shown [13].

While the abrupt, simultaneous transitions in both M and ρ shown in Fig.3a suggest a strong spin-charge coupling when B||*a*-axis, it is clear that the spin-polarized state can at most lower the resisitivity by one order of magnitude evident in both Figs. 2a and 3a. Then, *what is the origin that reduces the resistivity by three orders of magnitude when B||b-axis and when the spin-polarized state is destabilized* (see the inset in Fig3*)*? It is this issue that reflects the physics fundamentally different from that driving other magnetoresistive materials including the manganites where a spin-polarized state is essential for CMR [2].

It becomes increasingly clear that the novel behavior observed in the ruthenate is predominantly associated with the role of the orbital degree of freedom and its coupling to the spin and lattice degrees of freedom. The orbital degeneracy can be lifted by both the Jahn-Teller distortion and the spin-orbit coupling. As reported earlier, the abrupt decrease in the *c*-axis lattice parameter at $T_{MI}$ suggests the Jahn-Teller distortion that lifts the degeneracy of the $t_{2g}$ orbitals by lowering the energy of the $d_{xy}$ orbital relative to that of the $d_{yz}$ and $d_{xz}$ orbitals [3,5] and facilitates the orbital ordering. On the other hand, the orbital degeneracy can be also lifted by the spin-orbit interaction, which is accompanied by simultaneous spin symmetry breaking evidenced by the occurrence of the magnetic ordering at $T_{MI}$ in M for the *a*-axis (see the inset in Fig.1). This anisotropic magnetic behavior illustrated in Figs.1-3 can be attributed to the existence of the orbital ordering along the *a*-axis below $T_{MI}$ that is stabilized by the spin-polarized state via the spin-orbit interaction. Loosely speaking, spin ferromagnetism must be accompanied by antiferrorbital order and vise versa. Therefore, applying B along the *a*-axis in effect strengthens the orbital-ordered state by inducing the ferromagnetism. The magnetoresistive behavior for B||*a*-axis shown in Fig.2a and Fig.3a is thus due entirely to the spin-polarization or reduction of spin scattering, while the still nonmetallic behavior arises from the orbital ordering that forbids neighbor hopping between orbitals. Conversely, applying B along the *b*-axis effectively destabilize the ferromagnetic state, thus the orbital-ordered state. The field-tuned collapse of the orbital-ordered state systematically and drastically increases the electron hopping amplitude or kinetic energy, leading to the coloassal magnetoresistvity as it is unequivocally illustrated in Fig.2b and Fig.3. In line with the argument, it is worthy pointing out that the magnetic easy-axis starts to rotate away from the *a*-axis near $T_{MI}$ and becomes somewhat parallel to the b-axis at $T_N$ with $M_s$ being only 0.8 $\mu_B$/Ru (see the inset in Fig.3) [4,6]. This rotation destabilize the spin-orbit coupling favorable for the orbital ordering, resulting in the rare antiferromagnetic metallic state seen intermediate between $T_{MI}$ and $T_N$ [6].

The orbital ordered state might also be manifested by a possible density wave below $T_{MI}$. Shown in Fig.4 is the resistivity for the *a*-axis, $\rho_a$, as a function of dc electric field, E, at B=7 T applied along the *a*-axis. The non-ohmic behavior above the threshold field, $E_T$, suggests sliding density wave transport. The onset of the nonlinear conduction is also evidenced by the current-voltage (I-V) characteristic shown in the inset. An "S " shaped nonlinear behavior is also observed at B=0, but the rapidly increasing resistivity with decreasing temperature below $T_{MI}$ creates ambiguity due to possible self-heating. As discussed above, applying B large than 6 T along the *a*-axis drastically reduces the magnitude of $\rho_a$ with the non-monotonous temperature dependence (see the inset 2). Should the non-linear behavior be due to the self-heating effect, E dependence of $\rho_a$ and the I-V curve would be characteristically different when measured below and above the peak at 25 K seen in $\rho_a$ (T) at 7 T shown in the inset 2. The formation of the density wave below $T_{MI}$, although subject to a thorough investigation [14], provides additional evidence for the existence of the orbital-ordered state brought about by the highly anisotropic spin-lattice-orbital coupling.

Recently, a theoretical study using the Hubbard model with Coulombic and phononic interactions predicts the existence of a ferromagnetic orbital ordered state and coloassal magnetoresistance in the ruthenates [13]. The phase diagram generated in this study shows that the ferromagnetic orbital ordered state is stabilized when both the inter-orbital Coulomb interaction and the phonon self-trapping energy are sufficiently strong [13]. The study also suggests possible colossal magnetoresistive behavior due to a strong competition between ferromagnetic and antiferromagnetic states. The general agreement between the theoretical and experimental results further validates the crucial role of the orbital ordering driving the novel phenomena.

In conclusion, the colossal magnetoresistance reported here is due to the collapse of the orbital ordering through destabilizing the spin-polarized state. The physics involved is fundamentally different from that governing all other magnetoresistive materials.

This work was supported by NSF grants DMR-0240813 and DMR-0100572. G.C. is grateful to Dr. Ganpathy Murthy for very helpful discussions.

14. We have searched unsuccessfully for narrow-band noise associated with density wave sliding, but will continue investigating the issue.

Captions:

Fig.1. Temperature dependence of magnetization, M, for the *a*- (a) and *b*-axis (b) for a few representative magnetic fields, B. Inset: M vs. T for the *a*- and *b*-axis at B=0.5 T.

Fig.2. Temperature dependence of the *c*-axis resistivity, $\rho_c$, at a few representative B applied along the *a*-axis (a) and the *b*-axis (b) for 1.2 K<T<80. Inset: $\rho_c$ vs. T for $B_{||a}$=28 T. The temperature dependence of the *a*-axis resistivity $\rho_a$ is identical to that of $\rho_c$ presented here, therefore not shown.

Fig.3. Field dependence of $\rho_c$ for B||*a*-axis (a) and B||*b*-axis (b) for a few representative temperatures from 0.6 K to 49 K. M at 5 K for the *a*-axis and the *b*-axis is shown in Fig.3a (right scale). Inset: M as a function of angle, Θ, between the *a*-axis (Θ=0) and B at B=6.2 T for temperatures indicated. Note that M decreases abruptly when B rotates away from the *a*-axis at T<40 K and that the easy-axis starts to rotate to the *b*-axis (Θ=90°) with increasing temperature when T>42 K. The field dependence of the *a*-axis resistivity $\rho_a$ is nearly identical to that of $\rho_c$ presented here.

Fig.4. Resistivity for the *a*-axis, $\rho_a$, as a function of dc electric field, E, at B=7 T applied along the *a*-axis for a few temperatures indicated. Inset 1: I-V curves for temperatures indicated. Inset 2: Temperature dependence of $\rho_a$ at B=7 applied along the *a*-axis. Note that $\rho_a$ shows a peak near 25 K. That the nonlinear transport behavior below and above 25 K is characteristically the same indicates the self heating effect, if any, is not dominant.

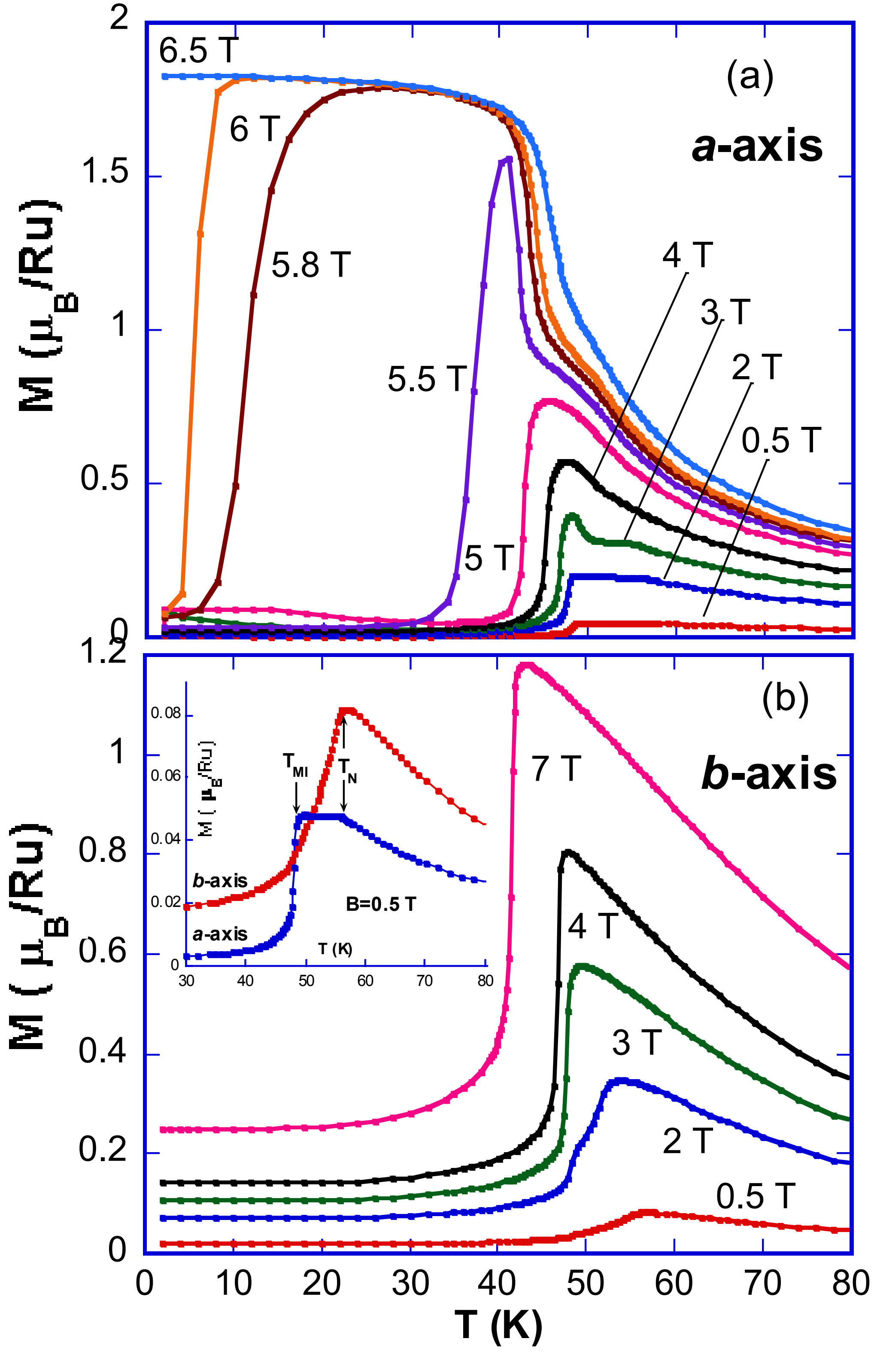

Fig.1 Cao

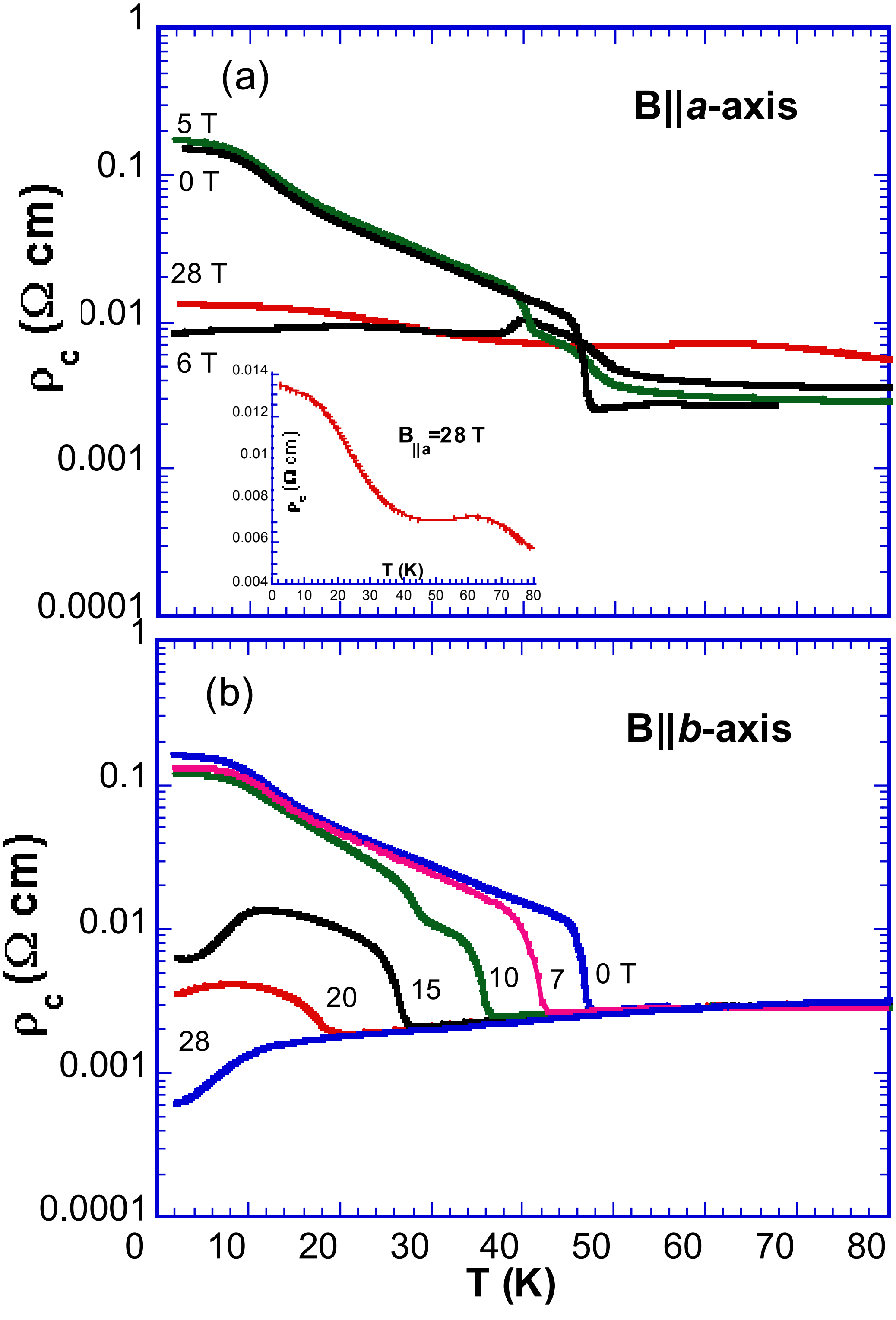

Fig.2 Cao

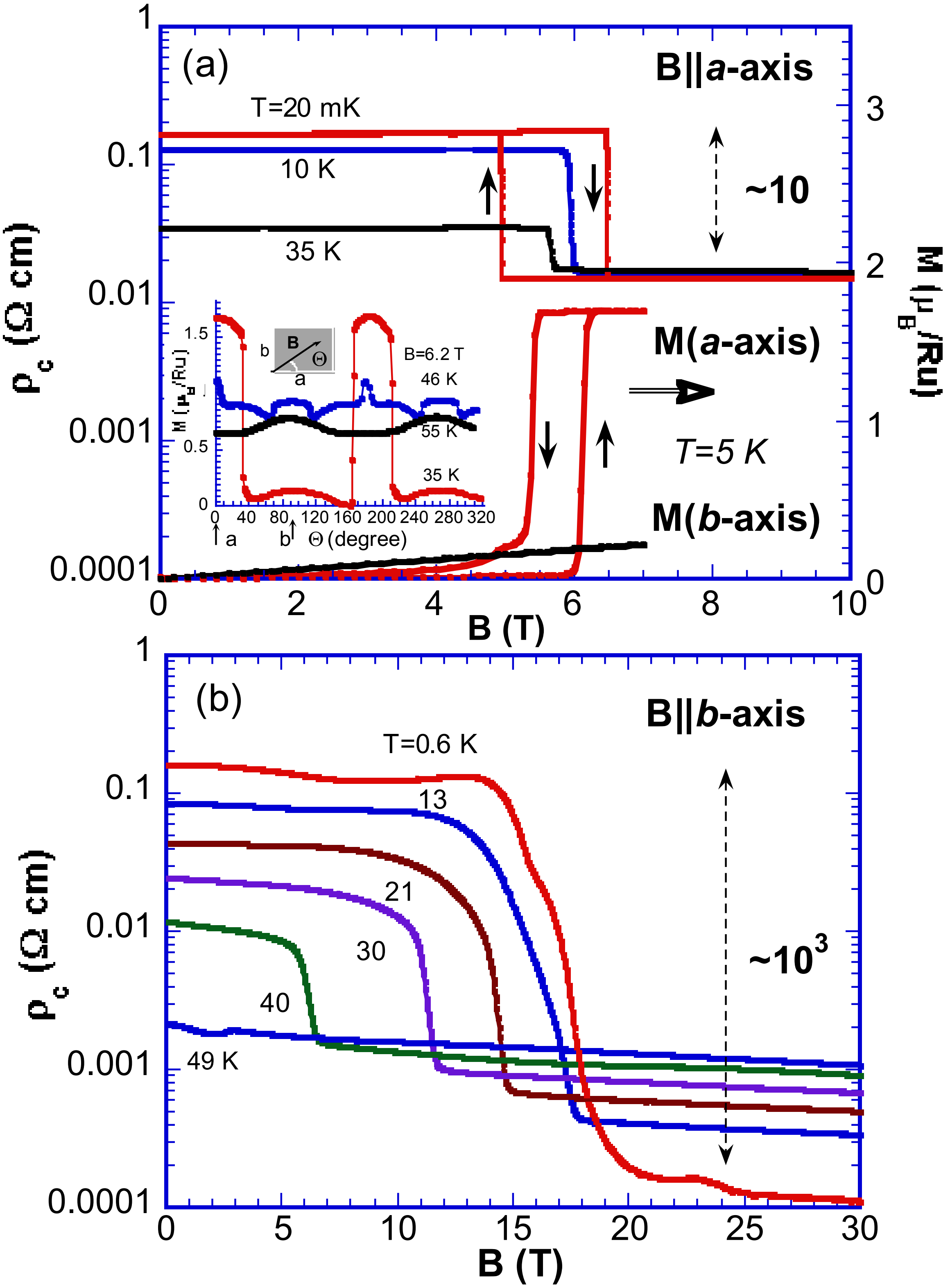

Fig.3 Cao

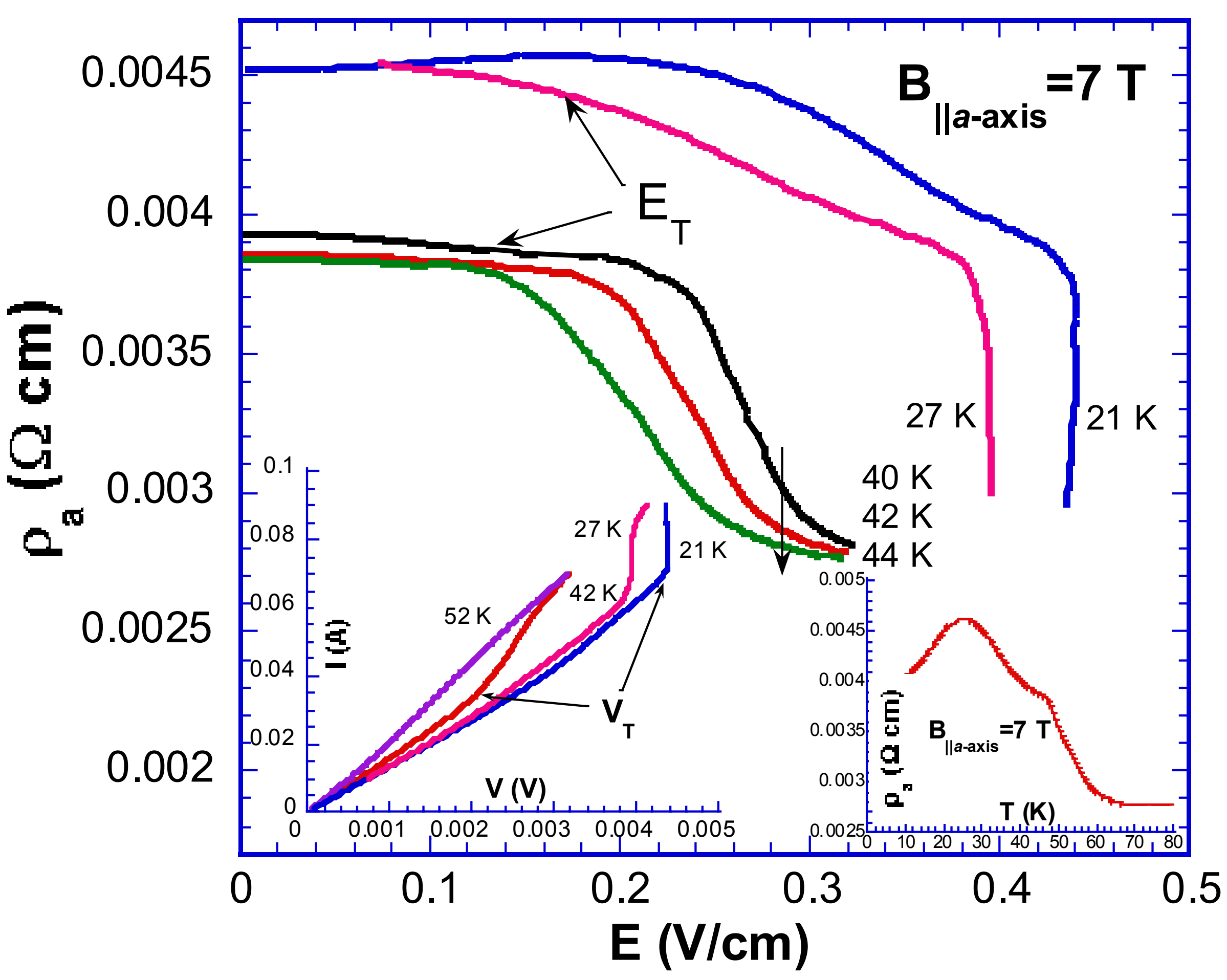

Fig.4 Cao